\documentclass[fleqn,twoside]{article}
\usepackage{espcrc2}
\usepackage{graphicx}

\def\NPB#1#2#3{{\it Nucl.~Phys.} {\bf{B#1}} (#2) #3}
\def\PLB#1#2#3{{\it Phys.~Lett.} {\bf{B#1}} (#2) #3}
\def\PRD#1#2#3{{\it Phys.~Rev.} {\bf{D#1}} (#2) #3}

\def\ZPC#1#2#3{{\it Z.~Phys.} {\bf C#1} (#2) #3}

\def\PR#1#2#3{{\it Phys.~Rep.} {\bf#1} (#2) #3}

\def\be{\begin{equation}}
\def\ee{\end{equation}}
\def\bea{\begin{eqnarray}}
\def\eea{\end{eqnarray}}
\def\ba{\begin{eqnarray}}
\def\ea{\end{eqnarray}}
\def\beq{\begin{equation}}
\def\eeq{\end{equation}}
\def\frac#1#2{{{#1}\over {#2}}}
\def\abs#1{\left|#1\right|}

\def\as{\alpha_s}

\newcommand{\AmS}{{\protect\the\textfont2
  A\kern-.1667em\lower.5ex\hbox{M}\kern-.125emS}}

\title{Polarized parton distributions from charged--current
deep--inelastic scattering and future neutrino factories}
       
\author{G. Ridolfi\address[MCSD]{INFN Sezione di Genova,\\
        Via Dodecaneso 33, I-16146 Genova, Italy}}
       
\begin{document}

\begin{abstract}
The capabilities of a neutrino factory in the determination of
polarized parton distributions from charged--current deep--inelastic
scattering experiments is discussed.  We present a study of the
accuracy in the determination of polarized parton distributions that
would be possible with such a facility.  We show that these measurements
have the potential to distinguish between different theoretical
scenarios for the proton spin structure.
\vspace{1pc}
\end{abstract}

% typeset front matter (including abstract)
\maketitle

Inclusive deep-inelastic scattering (DIS) experiments with polarized
beams~\cite{spinrev} only allow a measurement of C-even polarized
parton distributions:
\be
\Delta q^+=\Delta q+\Delta\bar{q}
\ee
In this context, the determination of individual flavors is difficult
because of the weak sensitivity to the polarized strange quark density
$\Delta s$; quark-antiquark separation, on the other hand, is entirely
impossible.  No polarized charged--current experiment has been
performed so far.  Given the intensity of presently available neutrino
beams, it would be impossible to build a polarized target of the
needed size.  Charged--current DIS with electron or muon beams, on the
other hand, would require a very high energy, since the exchange of
weak vector bosons is involved (this would be a possibility for HERA with
polarized beams).

An interesting possibility is offered by intense neutrino
beams~\cite{revnu} arising from the decays of muons along straight
sections of the accumulator of a muon storage ring. A facility of this
kind would allow an accurate decomposition of the partonic content of
the nucleon in terms of individual (spin-averaged and spin-dependent)
flavor densities.

The hadronic tensor relevant for charged--current polarized DIS can be
parametrized in terms of five polarized structure functions $g_i$,
$i=1,\ldots,5$ (see ref.~\cite{FMR} for details). Two of them ($g_2$
and $g_3$ in the conventions of ref.~\cite{FMR}) give no contribution
when the nucleon mass in neglected with respect to the hard scale
$Q^2$. Furthermore, $g_4$ and $g_5$ are not independent of each other
because of a Callan-Gross relation.  Therefore, there are only two
independent polarized structure functions at leading twist (we will
choose $g_1$ and $g_5$).

We assume that both $\nu$ and $\bar\nu$ beams will be available, and
therefore that all four structure functions $g_1^{W^\pm},g_5^{W^\pm}$
will be measured. Under this assumption, it can be shown~\cite{FMR}
that, taking suitable combinations of structure functions for
different targets, and combining measurements below and above the
charm threshold, a complete separation of all four flavors and
antiflavors is possible.

No extra information is provided by neutral--current structure function
measurements. However, they allow to perform consistency checks. In
particular, a direct measurement of the singlet quark combination
$\Delta\Sigma^+=\Delta u^+ +\Delta d^+ +\Delta s^+$ is an interesting
and independent check of present indirect information.  In fact, the
first moment of $\Delta\Sigma^+$ can be directly measured in
neutral--current DIS only assuming the knowledge of non-singlet first
moments from other experiments.  All other moments can only be
inferred from the study of scaling violations.

A next-to-leading order analysis in perturbative QCD is possible:
coefficient functions to order $\alpha_s$ and splitting function
to order $\alpha_s^2$ are known since long time~\cite{NLO}.
A full NLO evolution code for data analysis has been built~\cite{FMR},
based on previous work for neutral--current structure functions~\cite{ABFR}.

The C-odd combinations of polarized quark densities, 
\be
\Delta q^-=\Delta q-\Delta\bar{q},
\ee
are at present almost completely unknown. Some information on their
first moments is provided by semi-inclusive
experiments~\cite{semiinc}, which indicate that $\abs{\Delta\bar q}\ll
\abs{\Delta q^+}$ for the two light flavors $u$ and $d$.

Positivity of cross sections requires, at leading order,
\be
\abs{\Delta q}\leq q\quad\quad\abs{\Delta\bar q}\leq \bar q
\ee
for each individual flavor. Next-to-leading corrections to these
relations can be shown to be of the order of a few percent.  Using our
present knowledge on $\Delta q^+$, it can be shown that, for $q=u,d$,
the assumption $\Delta q^-=0$ (or $\Delta\bar q=\Delta q$) is
incompatible with the positivity bound in the large-$x$ region.  In
the case of polarized $s$, both $\Delta\bar s=0$ and $\Delta\bar
s=\Delta s$ are instead allowed.

A central issue in physics of polarized nucleons is to explain the
unexpected smallness of the axial charge 
\be
a_0(Q^2)=\int_0^1dx\,\left(\Delta\Sigma^+-\frac{n_f\as}{2\pi}\,\Delta g\right),
\ee
where $\Delta g$ is the polarized gluon distribution.
Present data indicate
that $a_0$ is compatible with zero, but values as large as
$a_0(10$~GeV$^2)=0.3$ are not excluded.  Different theoretical
scenarios have been proposed.  In one of them, it is assumed that a
cancellation between a large (scale--independent, or AB--scheme)
$\Delta\Sigma^+$ and a large $\Delta g$ takes
place. In this case, $\abs{\Delta u^+}, \abs{\Delta d^+} \gg
\abs{\Delta s^+}$ (in the AB--scheme), as expected in the quark model.
This is usually called the `anomaly' scenario.  Alternatively, one may
assume that $\Delta g$ is indeed small, and that $\Delta s^+$ is large
and negative.  This might be explained by invoking non--perturbative,
instanton--like vacuum configuration (`instanton' scenario). In this
case, $\Delta s =\Delta\bar s$.  In the `skyrmion' scenario, $\Delta
s^+$ large, but $\Delta s$ is significantly different from $\Delta
\bar s$.

Future experiments in polarized DIS will aim at establishing with the
best possible precision how small is the axial charge, and how large is
the polarized gluon distribution. Furthermore, it will be important to
understand whether $\Delta s$ is large, compared to the light flavor
distributions, and how different it is from $\Delta\bar s$.  In the
following we will explore the potential of a neutrino factory in
answering these questions. The parameters of a machine of this kind
are still under study; to be definite, we will consider the case of
a neutrino beam generated by a muon beam with energy
$E_{\mu}=50$~GeV and number of muon decays per year
$N_{\mu}=10^{20}$.

How accurately will charged--current structure functions be measured
at a $\nu$ factory? The various structure functions can be extracted
from the data exploiting the different $y$ dependence of each
component.  In this respect, the wide-band nature of the $\nu$ beam is
of great help. In fact, since
\be
y=\frac{Q^2}{2xmE_\nu},
\ee
one can have data points at different values of $y$ in the same
$x,Q^2$ bin, by varying $E_\nu$.  The measurement of $\mu$ and
hadronic recoil energies allows an event-by-event reconstruction of
$x$, $y$ and $Q^2$ (we adpot here the usual notation for DIS:
$\nu(k)+N(p)\to \mu(k')+X$, $Q^2=-q^2=-(k-k')^2$, $x=Q^2/(2pq)$,
$y=qk/pk$).

In order to estimate the impact of charged--current DIS data from a
$\nu$ factory on our knowledge of the proton spin structure,
we adopted the following procedure.
\begin{enumerate}
\item
We have computed the expected errors on the measurements of $g_1$ and $g_5$
in a region of the $x,Q^2$ plane compatible with the features of the device.
With the parameter choice of ref.~\cite{FMR}, this corresponds to
about $0.01 \leq x \leq 0.7$ and 
$1\,{\rm GeV}^2\leq Q^2 \leq 100\,{\rm GeV}^2$.

\item
We have performed new fits of existing neutral--current data. The
resulting values for 
\ba
&&\eta_\Sigma=\int_0^1 dx\,\left(\Delta u^+ +\Delta d^+ +\Delta s^+ \right)\\
&&\eta_3=\int_0^1 dx\,\left(\Delta u^+ -\Delta d^+ \right)\\
&&\eta_8=\int_0^1 dx\,\left(\Delta u^+ +\Delta d^+-2\Delta s^+ \right)\\
&&\eta_g=\int_0^1 dx\,\Delta g
\ea
and the axial charge $a_0$ at $Q^2=10$~GeV$^2$
are shown in table~\ref{fits}. Note that we have also performed a fit
with $\eta_g$, the first moment of the gluon distribution at $Q^2=1$~GeV$^2$,
forced to zero.
\begin{table*}[htb]
\caption{
Fits to neutral--current polarized DIS data. In the second column,
the first moment of the polarized gluon density at $Q^2=1$~GeV$^2$ is fixed to
zero.}
\label{fits}
\begin{tabular}{|c|c|c|} 
\hline
              & generic fit      & $\eta_g=0$ fit \\
\hline  
$\eta_\Sigma$ & $0.38\pm 0.03$   & $0.31 \pm 0.01$  \\
$\eta_g$      & $0.79\pm 0.19$   & $0$              \\
$\eta_3$      & $1.110\pm 0.043$ & $1.039\pm 0.029$ \\
$\eta_8$      & $0.579$          & $0.579$          \\
\hline
$\eta_u$      & $0.777$          & $ 0.719$         \\
$\eta_d$      & $-0.333$         & $-0.321$         \\
$\eta_s$      & $-0.067$         & $-0.090$         \\
\hline
$a_0$         & $0.183\pm 0.030$ & $0.284\pm 0.012$ \\
\hline
\end{tabular}
\end{table*}
This fit has a higher $\chi^2$, but once theoretical
uncertainties are taken into account a vanishing gluon distribution
can only be excluded at about two standard deviations. 
The last three rows in table~\ref{fits} show the values of
the first moments of $\Delta q^+$ at $Q^2=1$~GeV$^2$ for $q=u,d,s$.

\item
We have produced sets of ``fake'' data for charged--current structure functions
according to three different assumptions:
\begin{enumerate}
\item generic fit of table~\ref{fits} and $\Delta\bar s=0$
(anomaly scenario);
\item $\eta_g=0$ fit of table~\ref{fits} and $\Delta\bar s=\Delta s$ 
(instanton scenario);
\item $\eta_g=0$ fit of table~\ref{fits} and $\Delta\bar s=0$ 
(skyrmion scenario).
\end{enumerate}
In all three cases, we have fixed $\Delta\bar u=\Delta\bar d=0$,
consistently with the indications of semi-inclusive DIS data, and with
positivity constraints.  The fake data were then gaussianly
distributed about their central values, with the errors determined at
step 1. Data with $x>0.7$ and/or error larger than 50 were discarded
(the structure functions being typically of order 1). In this way, a
total of 8 sets of fake data ($g_1$ and $g_5$ with $\nu$ and $\bar\nu$
beams and proton or deuteron targets) were generated, with $\sim 70$
data points in each set (the neutral--current real data correspond to
176 points).

\item
We have performed a global fit of real and fake data. The results are shown in
table~\ref{refit}.
\begin{table*}[htb]
\caption{
Best--fit values of the first moments for the real and fake data and
fits discussed in the text. }
\label{refit}
\begin{tabular}{|c|c|c|c|} 
\hline
              &`anomaly' refit    & `instanton' refit & `skyrmion' refit\\
\hline
$\eta_\Sigma$ & $0.39  \pm 0.01$  & $0.321\pm 0.006$  & $0.324\pm 0.008$\\
$\eta_g$      & $0.86  \pm 0.10$  & $0.20 \pm 0.06$   & $0.24 \pm 0.08$ \\
$\eta_3$      & $1.097 \pm 0.006$ & $1.052\pm 0.013$  & $1.066\pm 0.014$\\
$\eta_8$      & $0.557 \pm 0.011$ & $0.572\pm 0.013$  & $0.580\pm 0.012$\\
\hline
$\eta_u$      & $ 0.764\pm 0.006$ & $0.722 \pm 0.010$ & $ 0.728\pm 0.009$\\
$\eta_d$      & $-0.320\pm 0.008$ & $-0.320\pm 0.009$ & $-0.325\pm 0.009$\\
$\eta_s$      & $-0.075\pm 0.008$ & $-0.007\pm 0.007$ & $-0.106\pm 0.008$\\
\hline
$a_0$         & $ 0.183\pm 0.013$ & $0.255\pm 0.006$  & $ 0.250\pm 0.007$\\
\hline
\end{tabular}
\end{table*}
The rows labelled $\eta_u$, $\eta_d$ and $\eta_s$ now give the
best--fit values and errors on the first moments of $\Delta q^-$ for
$u,d,s$. 
\end{enumerate}

Comparing the values of table~\ref{refit} with those of our original
fits leads to an assessment of the impact of charged--current data on
our knowledge of the polarized parton content of the nucleon.  As
announced, the precision on the singlet quark first moment is very
significantly improved by the charged--current data: the error on the
first moment of $\Delta\Sigma^+$ is now of a few percent in comparison
to about 10\% with neutral--current DIS.  With this accuracy, the
`anomaly' scenario can be experimentally distinguished from other
scenarios, at the level of several standard deviations.  
Correspondingly, the improvement in the precision on the gluon first
moment is sufficient to distinguish between the two scenarios.

The determination of the singlet axial charge is improved by an amount
comparable to the improvement in the determination of the singlet
quark first moment. It will be possible to establish at the level of a
few percent whether the axial charge differs from zero or not.  The
determination of the isotriplet axial charge is also significantly
improved: the improvement is comparable to that on the singlet quark.
This would allow an extremely precise test of the Bjorken sum rule,
and an accurate determination of the strong coupling.  Finally, the
octet C--even combination is now also determined with an uncertainty
of a few percent. Therefore, the strange C--even component can be
determined with an accuracy which is better than 10\%. Comparing this
direct determination of the octet axial charge to the value obtained
from baryon decays would allow a test of different existing models of
SU(3) violation.

Let us now consider the results for the first moments of the C--odd
distributions. We see that in the case of $u$ and $d$ quarks they can
be measured with an accuracy of a few percent, just sufficient to
establish whether the up and down antiquark distributions, which are
constrained by positivity to be quite small, differ from zero, and
whether they are equal to each other or not. Furthermore, the strange
C--odd component can be determined at a level of about 10\%, which is
enough to distinguish between the two cases $\Delta s^-\sim 0$ and
$\Delta s^-\sim \Delta s^+$.  The `instanton' and `skyrmion' scenarios
can thus also be distinguished at the level of several standard
deviations.

In conclusion, we have presented a review of the NLO formalism for
charged--current polarized DIS, and its numerical implementation. We
have described present experimental and theoretical constraints on the
C-odd polarized quark distributions, and we have explored the
capabilities of a neutrino factory in the study of polarized DIS. We
have shown that with such a facility the first moments of C--even
distributions can be measured with accuracies which are up to one
order of magnitude better than the current uncertainties.  This would
allow a definitive confirmation or refutation of the `anomaly'
scenario for the proton spin structure, compared to the `instanton' or
`skyrmion' scenarios. A test of models of SU(3) violation would also
be possible.

The first moments of the C--odd distributions $\Delta u^-,\Delta d^-$
can be determined at the level of few percent, while $\Delta s^-$
could only be measured at the 10\% level.  This would be sufficient to
test for instrinsic strangeness, and distinguish between `skyrmion'
and `instanton' scenarios.

Determining the shapes of the distributions is a more difficult
task. The precision on the shapes of the combinations of structure
functions that correspond to individual flavor distributions at
leading order is reasonably good, but their determination beyond
leading order is severely limited by the uncertainty on the shape of
the gluon distribution. It turns out that $\Delta u(x)$ and $\Delta
d(x)$ could only be measured with a precision around 15-20\%, while no
significant shape information can be obtained for $\Delta s(x)$.

\section*{Acknowledgements}
I wish to thank S. Forte for interesting
discussions on the subject of this talk.

\end{document}